\documentclass{ws-procs961x669}            % default, citations in superscript
\begin{document}
	\newcommand {\be}{\begin{equation}}
	\newcommand {\ee}{\end{equation}}
	\newcommand {\ba}{\begin{eqnarray}}
	\newcommand {\ea}{\end{eqnarray}}
\title{A dark matter solution for the XENON1T electron excess and the galactic center 511 keV line }

\author{Yasaman Farzan }

\address{School of physics, Institute for Research in Fundamental Sciences (IPM)\\P.O.Box 19395-5531, Tehran, Iran\\
$^*$E-mail: yasaman@theory.ipm.ac.ir\\
http://physics.ipm.ac.ir/}

\begin{abstract}
The excess of the 511 keV line from the Milky Way galactic bulge, confirmed by the INTEGRAL detector, is a longstanding mystery. The morphology of the line appears  to be proportional to the square of the dark matter density, hinting towards a dark matter origin. On the other hand, in 2020, XENON1T has reported an excess of electrons with a recoil energy of $2-4$ keV. We present a model based on a dark matter of a few MeV mass that decays into a pair of pico-charged particles with a lifetime much larger than the age of the Universe. The magnetic field of the galaxy accumulates these relativistic pico-charged particles whose scattering on the electrons can explain the signal reported by XENON1T. The  annihilation of the pico-charged particles in the galactic bulge leads to $e^-e^+$ production and therefore to an excess of the 511 keV line. We review the present observational bounds and the strategies to test the model.
\end{abstract}


\bodymatter

\section{Introduction}\label{aba:sec1}

Various observations such as galactic rotation
curves, structure formation, dynamics of the galaxy clusters, collision of bullet clusters and anisotropies of Cosmic Mircrowave Background (CMB) point towards existence of a new form of matter all over the universe known  as Dark Matter (DM). The constituent particles  of DM should be either electrically neutral or their electric charge should be much smaller than the electron electric charge. Otherwise, they will emit photons and will not remain dark. The dark matter lifetime should be much longer than the age of the Universe. The observation of bullet clusters constrain the self-interaction of the dark matter particles indicating that they cannot have strong interactions. Last but not least, at the onset of structure formation, the dark matter should be non-relativistic. None of the  Standard Model (SM) elementary particles satisfies these conditions.  As a result, the observational hints for dark matter opens a window towards new physics beyond the standard model by introducing a new neutral (meta)stable massive particle that can play the role of dark matter. \footnote{The primordial blackholes as dark matter candidates have received popularity after observation of gravitational waves by LIGO and VIRGO. We will not however pursue this idea in this letter.}

The models for dark matter candidates are quite diverse. Depending on the details of the model, different strategies for dark matter detection have been developed. In a wide class of models known as WIMP, where dark matter has a mass in the wide range of MeV to 100 TeV and it weakly interacts with the Standard Model (SM) particles \footnote{Here, by weak interactions, we do not necessarily mean nuclear weak interaction, characterized by the SM $W$ and $Z$ gauge boson exchange.},  there are three approaches to dark matter detection: (1) direct dark matter detection, (2) indirect dark matter detection and (3) dark matter production in colliders.
No conclusive evidence for dark matter detection has been reported from any of these experiments so far but every now and then, a signal is reported by indirect search experiments which may be interpreted as dark matter. Some of these signals disappear by collecting more data or are eventually proved to be artifacts. There are however signals that pass the test of time  but after some period of time simply go out of fashion because the simplistic dark matter models developed to explain them become ruled out by independent observations. 
However, the dark sector does not need to be simplistic. Indeed, judging from the rich and dazzling structure of the SM, simplicity and minimalism is not the taste of the physics of the  elementary particles. Going beyond simplistic dark matter models, it may be possible to resurrect dark matter explanation for the signal.

One example of the signals that has survived the test of time is the 511 keV line observed  from galactic center which must come from the positronium (the $e^-e^+$ atom) decay: $e^-+e^+\to \gamma+\gamma$. The intensity of the line indicates an excess of $e^+$ in the galactic center. Although the excess may come from the SM sources such as pulsars but entertaining a DM origin is too tantalizing to dismiss, especially that the morphology of the intensity of the line follows a $\rho_{DM}^2$ distribution, further pointing towards DM pair annihilation as the origin of the excess.
In Ref.~\cite{Boehm:2003bt}, it had been shown that dark matter  with mass of a few MeV can explain the excess of the 511 keV line observed from the galactic center by pair annihilation into pairs of electron positron with a cross section of $\langle \sigma({\rm DM+DM}\to e^-e^+)v\rangle \sim O(10^{-38})$ cm$^2$ . This model implies huge energy pump into the plasma at the era of recombination when the DM density is high, through annihilation into $e^-e^+$. Such entropy pump can distort the CMB fluctuation pattern. The Planck data therefore rules this solution out
\cite{Wilkinson:2016gsy}.
Moreover, this model is disfavored by the lack of observation of the  511 keV line from dwarf galaxies \cite{Sieg}.

Ref.~\cite{Farzan:2017hol} proposes a dark matter model for the 511 keV line from the center of the Milky Way (MW) that can avoid the bounds from CMB and explain the lack of the 511 keV line from satellite dwarf galaxies. In this model, dark matter is composed of a scalar particle, $X$, with a mass of a few MeV which can be identified with the SLIM particle \cite{SLIM,slimMODEL}. $X$ decays into a pair of pico-charged particle ($C+\bar{C}$) with a lifetime much longer than the age of Universe: $\Gamma(X\to C+\bar{C} )t_0 \ll 1$ in which $t_0$ is the age of the Universe.
The magnetic field of the MW accumulates the $C$ and $\bar{C}$ particles. The $C\bar{C}$ pairs eventually annihilate with each other producing $e^-e^+$ pairs. At the recombination and during the dark ages, the density of $C\bar{C}$ would be too small to pump energy to the plasma so the CMB bound can be avoided. Moreover, the magnetic field in the dwarf galaxies may be too weak to accumulate $C\bar{C}$ so the lack of the 511 keV line from them can be explained.

Ref.~\cite{Farzan:2017hol} predicted a signal at direct dark matter search experiments with low energy threshold. The relativistic $C$ particles wandering around can impart a recoil energy of a few keV via Coulomb interaction to the electrons. In 2020, XENON1T reported an excess of electrons with recoil energy of 2--4 keV \cite{Aprile:2020tmw}.  In Ref.~\cite{Farzan:2020llg}, we discussed the parameter space where the model can explain both signals. Refs.~\cite{DM+XENON} have also attempted to explain the XENON1T signal via DM.

This paper is organized as follows. In Sect.~\ref{model}, we describe
the model and the bounds that constrain its parameter space. In Sect.~\ref{511}, we show how
the model can explain the 511 keV line from the galactic bulge. In Sect.~\ref{XENON1t}, we describe  how the model can explain the XENON1T data. We show that the XENON1T electron recoil spectrum can also set the strongest upper bound on $f=\Gamma(X\to C\bar{C})t_0$. A summary is given in Sect.~\ref{sum}. In Sect.~\ref{sum}, we also  discuss how the model can be tested with further
collection of the 511 keV data  from satellite  
galaxies and studying the correlation of the intensity with the magnetic field of the host [dwarf] galaxy. We also discuss the signatures in  diffuse gamma ray background from the inflight annihilation of positron. We also comment on the positron flux to be detected by Voyager.

\section{The model and its constraints \label{model}}
As mentioned in the introduction, the model is based on a scalar dark matter, $X$ with a mass of a few MeV which decays into a pair of pico-charged particles, $C\bar{C}$. The $X$ particle can be identified with the SLIM particle introduced in Ref.~\cite{SLIM}. In the  SLIM scenario, the DM is thermally produced via interaction with neutrinos and its abundance is set {\it a la} freeze-out scenario through lepton number violating annihilation into neutrino or antineutrino pair, $\langle \sigma(X+X\to \nu+\nu)v\rangle=\langle \sigma(X+X\to \bar{\nu}+\bar{\nu})v\rangle\sim pb$.
Within the SLIM scenario, the annihilation to neutrino or antineutrino pair takes place via the $t$-channel exchange of a new Majorana neutrino which has a Yukawa coupling with the SM neutrino and $X$. A $Z_2$ symmetry, under which only the new Majorana neutrino and $X$ are odd, guarantees the stability of $X$ which is the lightest among the $Z_2$ odd particles. The same $Z_2$ symmetry forbids Yukawa coupling between the leptons, the Higgs and the new Majorana fermions and as a result, neutrinos cannot obtain a Dirac mass at the tree level. The SM neutrinos obtain  a Majorana mass at one loop level within the SLIM scenario. The smallness of the neutrino mass is explained, thanks to the loop suppression despite a relatively large Yukawa coupling between, $X$ and neutrinos of order of $10^{-3}$. The model is testable by searching for the three body decay of Kaon and pion where along with the charged leptons, these new particles are emitted and appear as missing energy \cite{Farzan:2010wh}. Ref.~\cite{slimMODEL} shows how the SLIM scenario can be embedded within a UV complete model. Combining the condition $\langle \sigma(X+X) v\rangle \sim pb$ (to obtain the required DM relic abundance via the freeze-out mechanism) with the one-loop contribution to the neutrino mass to be equal to $\sim \sqrt{\Delta m_{atm}^2}$ sets an upper bound of $\sim 10$ MeV on the dark matter mass. On the other hand, $X$, being in thermal equilibrium with neutrinos at the time of the neutrino decoupling, cannot be lighter than a few MeV \cite{Sabti:2019mhn};
otherwise, it will lead to too many extra relativistic degrees of freedom.
If $X$ is not the SLIM and its production is not thermal, these lower and upper bounds on its mass can be relaxed. 

Taking the pico-charged particles $C$ and $\bar{C}$ to be scalars, the decay $X\to C\bar{C}$ can take place via a soft $Z_2$ symmetry breaking trilinear term of form $AX \bar{C} C$ with a very small $A$ making $X$ metastable.
Let us now discuss how $C$ and $\bar{C}$ acquire their minuscule electric charge. For this purpose, we need a new Abelian gauge symmetry such that 
the $C$ and $\bar{C}$ are charged under the new $U_X(1)$ gauge symmetry. We denote the new gauge boson with $A_\mu$. The kinetic and mass mixings between $A_\mu$ and the SM hypercharge gauge boson $B_\mu$ lead to a small electric charge for the $C$ and $\bar{C}$ particles. We follow the notation of Ref.~\cite{Feldman:2007wj} and write \be -\frac{A_{\mu \nu}A^{\mu \nu}}{4}-\frac{B_{\mu \nu}B^{\mu \nu}}{4}-\frac{\delta}{2}A_{\mu \nu}B^{\mu \nu}-\frac{1}{2}\left(\partial_\mu \sigma +M_1 A_\mu +\epsilon M_1 B_\mu\right)^2 \ , \ee
where $\delta, \epsilon\ll 1$.
In the  kinetic and mass basis, the three neutral gauge bosons are the SM $\gamma$, $Z$ bosons and a new gauge boson  $A'_\mu$ which we shall  call dark photon. To the first approximation in $\epsilon$ and $\delta$, the mass of $A'$ is decoupled from  $m_Z$ and is equal to $M_1$. Up to $O(\epsilon^2,\delta^2)$, the  coupling between the SM charged fermions, $f$ and $A'$ can be written as the following:
\be q' \bar{f}\gamma^\mu f A'_\mu  \ \ \ \ \ {\rm where} \ \ \ \ \  q'=e \cos \theta_W (\epsilon-\delta)Q_f, \label{e-d} \ee in
which $\theta_W$ is the weak (Weinberg) mixing angle.
The coupling of $C\bar{C}$ to  $A'$ is given by 
$g_X J_C^\mu A'_\mu$ where $J_C^\mu$ is the current of the $C$ particles.
Then, the $C$ particle obtains an electric charge of 
$$ q_C=-g_X \epsilon \cos \theta_W.$$  
From Eq.~(\ref{e-d}), we observe that in the limit $\epsilon \to \delta$, the SM fermions do not couple to $A'$. In this limit, if $A'$ is lighter than the $C\bar{C}$ pair with a mass of order of a few keV, it will become metastable with a lifetime much greater than the age of the Universe
\cite{Farzan:2017hol}. This means the $A'$  particles produced in the early universe remain as relics today and contribute as a subdominant DM component. In Ref.~\cite{Farzan:2020llg}, we do not however consider this limit and allow for a faster decay of $A'$. As shown in  Ref.~\cite{Farzan:2020llg}, the $U_X(1)$ symmetry can be identified with the $L_\mu-L_\tau$ symmetry with a gauge coupling of $g_{\tau-\mu}$ which opens up the possibility of fast decay $A' \to \bar{\nu}_\mu \nu_\mu,
\bar{\nu}_\tau \nu_\tau$. The decay can  relax the bound from supernova cooling, opening up the window $m_{A'}\sim 100$ keV, $q'\sim 10^{-10}-10^{-9}$ and $g_{\tau-\mu} \sim 10^{-9}$ . Such values of coupling are too small to produce $A'$ with significant abundance in the early universe so the bounds on the extra relativistic degrees of freedom from CMB and BBN can be relaxed. On the other hand, the $A'$ particles produced in the supernova core  decay into neutrino and antineutrino pairs in the outer layers of supernova. This process can severely change the supernova evolution. It can even facilitate  the shock revival.
Thus, in future with better observation of  supernovae and more advanced simulations, this part of the parameter range of the model can be tested. For the time being, however, this range of the parameter space is accepted within the uncertainties. The $A'$ particles escaping the core decay into $\nu \bar{\nu}$ pairs so similarly to the standard picture, the binding energy of the star will be emitted in the form of neutrinos to be observed by  detectors on the Earth such as Super-Kamiokande.

The $C$ particles produced by the $X$ decay will be relativistic at the production with a velocity far exceeding the escape velocity from the galaxy. They are kept in the galactic disc by the magnetic field. To accumulate the $C$ and $\bar{C}$ particles, their Larmour radius, $$r_c=(100~{\rm pc })\frac{3\times 10^{-11}}{q_C} \frac{m_X}{10~{\rm MeV}}\frac{1~ \mu{\rm Gauss}}{B}$$ should be smaller than the thickness of the galactic disc. This requirement sets a lower bound on $q_C$: $q_C>10^{-11}$, see Refs.~\cite{Farzan:2017hol,Farzan:2020llg}. The strongest upper bound on $q_C$ in this mass range comes from supernova cooling consideration and is $q_C<10^{-9}$, see Ref.~\cite{Davidson:2000hf}.

 As shown in Ref.~\cite{Chuzhoy:2008zy},
expanding supernova remnants can pump energy to the charged particles, enhancing their Larmor radius and expelling them from the galactic disc with a rate  of $(100~{\rm M}yr)^{-1}$ unless there is an efficient mechanism for energy loss for the charged particles.  As discussed in Ref.~\cite{Farzan:2017hol}, in the limit $\delta \to \epsilon$ where $A'$ is (meta)stable,  the relic background of $A'$ can behave as a coolant.  That is the $C$ and $\bar{C}$ particles can scatter off the background $A'$, losing energy. As discussed in Ref.~\cite{Farzan:2020llg}, in general case $\delta \ne \epsilon$ where there is no $A'$ background, another coolant should be introduced.  Following Ref.~\cite{Farzan:2020llg}, we denote the coolant with $Y$ and take  $10~{\rm eV}<m_Y<10~{\rm keV}$, $n_Y|_{local}=(\rho_{DM}|_{local}/\langle \rho_{DM}\rangle ) \langle n_Y\rangle$ with a coupling of $g_Y$ to the $C$ particles leading to the  scattering cross section of  $\sigma_S\sim g_Y^4/(4\pi E_C^2)$. The energy loss at each collision can be written as $ \Delta E_C=m_Yv^2 ({E_C}/{m_C})^2 .$

The energy of the $C$ particle at the production from $X\to C\bar{C}$ is equal to $m_X/2$. In the absence of energy pump from supernova, the time scale for the energy loss from $m_X/2$ to $m_C(1+v_f^2/2)$ is
\be \tau_E=\int_{m_C(1+v_f^2/2)}^{m_X/2} \frac{dE_C}{\Delta E_C}\frac{1}{\sigma_S vn_Y}\sim \frac{4\pi m_C^3}{g_Y^4 n_Ym_Y}\left( \frac{1}{v_f}-\frac{1}{v_i}\right). \ee
Equating $\tau_E$ with the time scale of the energy gain from expanding supernova remnants ({\it i.e.,} 100 Myr), we find
\be v_f =0.08 \left(\frac{0.25}{g_Y}\right)^4 \left(\frac{m_C}{3 {\rm MeV}}\right)^3 \frac{0.1 \times \rho_{DM}|_{local}}{n_Ym_Y}.\label{vf} \ee

Let us now discuss the bounds on $\Gamma_X=\Gamma(X\to C\bar{C})$ or equivalently on $f=\Gamma_X t_0$.  A rather strong  bound comes from the requirement that annihilation $C\bar{C} \to A'A'$ does not deplete the $C$ abundance of  the galactic disc: That is $\sigma(C\bar{C}\to A'A')n_Cv_ft_0<1$ which leads to $$f<10^{-2}\frac{m_X}{10~{\rm MeV}}\left(\frac{0.15}{g_X}\right)^4 \left(\frac{m_C}{5~{\rm MeV}}\right)^2 \frac{0.06}{v_f}.$$ A stronger bound comes from the requirement that the process of cooling during the history of the galaxy does not repel all the coolants. We should bear in mind that the coolant  is a subdominant DM component which contributes  less than 10 \% to the whole  dark matter content. Ref.~\cite{Farzan:2020llg} shows that this requirement implies $$f<10^{-3}\frac{m_X}{10~{\rm MeV}}\frac{m_Y}{10~{\rm keV}}\frac{10 \rho_Y}{\rho_{DM}}.$$
As we shall see in sect. \ref{XENON1t}, the bound from XENON1T detector on $f$ is much stronger.

In order to produce $e^-e^+$ from $C\bar{C}$ annihilation yet another interaction is required. We will discuss this coupling in Sect.~\ref{511}.
\section{ Explaining the 511 keV line \label{511}}

 The 511 keV photon line has been detected from the galactic bulge  of  the MW for more than 40 years \cite{Siegert:2015knp}. The INTErnational Gamma-Ray Astrophysics Laboratory (INTEGRAL) is a space telescope launched in 2002.  The  data from the SPI  spectrometer at INTEGRAL over the years has confirmed the 511 keV signal at 56 $\sigma$ C.L. The narrow photon line peaked at 511 keV, whose energy coincides with the rest mass of electron or positron, should come  from the annihilation of the electron positron pairs at rest which form  positronium atoms. The source of such large  positron abundance   in the bulge of MW is subject to debate. An intriguing possibility  can be  annihilation of DM leading to the $e^-e^+$ production.  Astrophysical  sources such as  pulsars  or $X$-ray binaries  may  also contribute to the positron flux but their characteristic spectral shapes do not fit  the observations \cite{Bandyopadhyay:2008ts}. Moreover, the   morphology  of the observed  $511$ keV line is better compatible with the $\rho_{DM}^2$ ditribution than what is predicted from these point sources, hinting further toward a DM annihilation origin \cite{ast}.

Ref.~\cite{Boehm:2003bt} proposes  DM  with mass of few MeV annihilating to $e^-e^+$ as a solution to the 511 keV line. With an annihilation  $\langle \sigma(DM+DM\to e^-+e^+)v \rangle\sim 10^{-2}$~pb, MeVish DM can account for the intensity  of the 511 keV line.  However, if the annihilation to $e^-e^+$ is
a $S$-wave  process, the reioinzing energy dump at recombination will cause delayed recombination which is in tension with the  CMB data \cite{Wilkinson:2016gsy}. If the annihilation is $P$-wave, at the freeze-out temperatures, the annihilation cross section will be   $\sim$10 pb, washing out the relic density of DM. Thus, we should go beyond this minimal scenario. In the following, we show that the model described in sect. \ref{model} can provide a reliable explanation.

The  $X$  DM particles  in the galaxy decay into $C\bar{C}$ with a  rate of $\Gamma_X$. If these $C$ particles are accumulated close to their production point, their number density after  $t^0$ will be 
\be n_C=\frac{\rho_X}{m_X} f \ \  \ \ {\rm with} \ \ \ \ f=\Gamma_X t^0\ee in which $f$ is the fraction of $X$ particles that have decayed.
According to the INTEGRAL findings \cite{Siegert:2015knp}, the  intensity of $511$ keV line from galactic center  is  $  \Phi_{511} = (0.96 \pm 0.07)\times 10^{-3} \rm{ph} ~\rm{cm^{-2}}~\rm{sec^{-1}}$. 
If each  $C\bar{C}$  annihilation  produces $N_{pair}$ pairs of $e^-e^+$ and subsequently $N_{pair}$ pairs of photons,
the flux from the galaxy bulge will be
\begin{equation}\label{511}
\Phi _{511} \simeq 2  N_{pair} \frac{\int _0 ^{r_b} (n_X(r)) ^2 f^2  \langle \sigma _{C \bar{C}} v \rangle 4 \pi r^2 dr}{4 \pi r_{sol}^2}
\end{equation}
where $r_{sol} \simeq 8$ kpc is the distance between us and  the  center of MW, $r_b\simeq 1$ kpc is the galactic bulge radius and  $n_X ={\rho_X (r)}/{m_X}$. 
In the minimalistic dark matter scenario in which the dark matter pair annihilates directly into $e^-e^+$, the flux is given by the formula in Eq.~(\ref{511}), replacing $N_{pair}f^2$ with 1 and $\sigma_{C\bar{C}}$ with $\sigma ({\rm DM+DM}\to e^-e^+)$. In this scenario, the cross section should be of order of $10^{-39}-10^{-38}$ cm$^2$ to account for the 511 keV line intensity which implies that in our model:
%Taking an NFW halo profile   and a local dark matter density of  0.4 ${\rm GeV/ cm^3}$ \cite{Hooper:2016ggc}, the $C\bar{C}$ annihilation cross section should be
\be \sigma_{C\bar{C}}\sim 2~nb\left(\frac{1.5\times 10^{-4}}{f}\right)^2 \left(\frac{m_X}{5~{\rm MeV}}\right)^2\frac{1}{N_{pair}} .\ee

%\begin{equation}
%\rho(r) = \rho_0 (\frac{R_s}{r})^\gamma \times \frac{1}{(1+ \frac{r}{R_s})^{3-\gamma}}
%\end{equation}
%where $\gamma = 1$ corresponds to the case of a standard NFW profile, $\rho_0$ is local dark matter density and $R_s$ is scale radius. Taking standard NFW profile, for  $\rho_0 = 0.3 %~\rm{GeV}~\rm{cm^{-3}}$ and $R_s = 20$ kpc \cite{Hooper:2016ggc}, expected flux would be given by
%\begin{equation}
%\Phi = 0.97 \times 10^{-3} ~\rm{cm^{-2}}~\rm{sec^{-1}} \left( \frac{\rho_0}{0.31 ~\rm{GeV}~\rm{cm^{-3}}} \right)^2 \left(  \frac{5 ~\rm{MeV}}{m_X} \right)^2 \left( \frac{f}{ 1.5 \times 10^{-4}}  %\right)^2 \left( \frac{\langle \sigma v \rangle}{100 ~\rm{pb}} \right).
%\end{equation}
%which is very close to the observed flux. Taking Einasto or non standard NFW halo profile also leads to similar flux.

Let us now discuss the couplings that leads to 
 the  $e^-e^+$  production from the $C\bar{C}$ annihilation. Taking effective coupling $(\bar{e}e)(\bar{C} C)/\Lambda$,  the cross section of $2$ nb implies $\Lambda < 100$ GeV which would be within the reach of the LEP collider and other accelerator experiments. We therefore invoke the possibility of intermediate light neutral particles, $\phi$, to avoid the bounds from the LEP. Then, the production of $e^-e^+$ takes place in two steps: First,  $C\bar{C}\to\phi \bar{\phi}$ and subsequently $\phi \to e^-e^+$. As a result,  $N_{pair}=2$.
The $\phi \to e^-e^+$ decay can take place through  \be g_\phi \phi \bar{e}e.\ee   As long as $g_\phi <10^{-11}$, the $\phi$ production   in the supernova core will be negligible. We need $g_\phi>10^{-15}$ in order for  the $\phi$  decay length to be shorter than 10~pc.  If  $\phi$ travels less than $\sim$100 pc  in the bulge (i.e., of order of the resolution of the INTEGRAL observatory) before decay, the $n_{DM}^2$ dependence of the 511 keV line morphology   will be maintained, as favored by observation. Thus, $0.3 \times 10^{-15}< g_\phi < 10^{-11}$.  $\phi$ is mainly  an electroweak singlet mixed  with the SM Higgs via  $$ a_\phi \phi |H|^2$$ Then, \be  g_\phi = \sqrt{2}\frac{a_\phi v}{m_h^2 }Y_e\ee in which $Y_e$ is the Yukawa coupling of the SM Higgs to the electron.  Fortunately,  the hierarchy between the $H$ and $\phi$ masses does not require  unnatural fine tuned cancellation at tree level because  $a_\phi\stackrel{<}{\sim}m_\phi$. In the early universe, $\phi$ can be produced via $e^-e^+ \to \phi \gamma$  with a number density $n_\phi \sim \langle \sigma_\phi v \rangle n_e^2 H^{-1} |_{T=m_\phi}$ where $\sigma_\phi=e^2 g_\phi ^2/8 \pi m_\phi^2$. Thus, during the period  between $T=m_\phi$ and  $\Gamma_\phi^{-1}$, we can write $n_\phi/n_\gamma=8 \times 10^{-8} (g_\phi/10^{-11})^2$.   As a result, $n_\phi$ at $T\sim 1$ MeV would be  small enough to satisfy the  BBN bounds. In the range of $g_\phi$ in which we are interested,  the decay of the $\phi$ particles takes place long before the recombination. For more details, see Ref.~\cite{Farzan:2017hol}.

\section{Direct detection of the $C$ particles in the galaxy\label{XENON1t}}
The $C$ particles, having a tiny electric charge, can have Coulomb interaction with the electrons and protons. It can also interact with matter fields by the $t$-channel exchange of dark photon, $A'$. The velocity of $C$  arriving at the detector is larger than the typical velocity of the DM particles ({\it i.e.,} velocity of $C$ $\gg 10^{-3}$). As a result, the recoil energy imparted by the $C$ collision can be much larger than that expected from the collision of a DM particle of mass of MeV. This rises the hope to search for $C$ particles by the direct dark matter search experiments, despite $C$ being light. In this section, we shall show how the electron excess with recoil energy of $2-4$ keV at XENON1T can be explained within our model. We will then show that XENON1T sets strongest upper bound on the fraction of the $X$ particle decaying into $C\bar{C}$, $f$.

According to Eq.~(\ref{vf}), the velocity of  $C$ particles around us is $v_f\sim 0.08$. The Larmor radius in  the Earth magnetic field  is $5\times 10^8~{\rm km}(10^{-11}/q_C)(m_C/3~{\rm MeV})$. Since the Larmor radius is much larger than the Earth radius,  the Earth magnetic field   cannot accumulate  the $C$ particles so their density  around the Earth will be similar to anywhere in the solar system (i.e., anywhere in a distance of 8 kpc from the MW galactic center): 
$(\rho_{DM}|_{local}/m_X) f$ in which $\rho_{DM}|_{local}\sim 0.4~
{\rm GeV}/{\rm cm}^3$.

 Consider a  $C$ particle with a velocity of $v_f$ scattering off a non-relativistic particle of mass, $m$. Up to corrections of $O(v_f^4)$, the maximum recoil energy, which corresponds to the backward scattering, can be written as
 \be  E_{max} =2m v_f^2 \frac{m_C^2}{(m+m_C)^2} .\label{max}
 \ee
For  $m_C\sim 1-5$ MeV, $v_f =0.08$  and $m=m_e$,  $ E_{max} =3-5.5$~keV which is tantalizingly   within the range of electron recoil excess observed by XENON1T \cite{Aprile:2020tmw}. Taking $m$ equal to the mass of  Xenon or of Argon,  $ E_{max} $ turns out to be much below the detection energy threshold of the experiments such as XENON1T \cite{Aprile:2018dbl}, DarkSide \cite{DarkSide-50} and even 
 the CRESST detector \cite{Schieck:2016nrp}. 
 
  The spectrum of recoiled electrons per detector  mass will be 
  $$\frac{dN}{dE_r}=\frac{Z_{out}}{m_N}(2 f\frac{\rho_X}{m_X}) \int f_C (v) \frac{d\sigma}{d E_r}vdv$$
  where $m_N$ is the mass of the nuclei inside the detector; {\it i.e.,}  $m_N=131$~GeV for the Xenon. $Z_{out}$ denotes  the number of the  electrons in the outer orbitals of the detector atoms whose  binding energies are smaller than $E_r$. Ref.~\cite{Hsieh:2019hug} shows that to a reasonable approximation,  the binding energy can be treated with a    step function in computing the scattering cross section. For Xenon, $Z_{out}=44$ which is the number of the electrons in the $n=3,4,5$ orbitals. These  outer electrons have velocities smaller than $v_f$ so we neglect their velocity in our analysis.   The velocity distributions of the $C$ and $\bar{C}$ particles  are given by $f_C(v)$. In our computation, we take 
  $f_C(v)=\delta( v-v_f)$ for simplicity.  As far as $|\epsilon|/2m_eE_r\gg |\epsilon-\delta|/(2m_eE_r+m_{A'}^2)$, the $t$-channel photon exchange gives the dominant contribution to the $C$ particles scattering off  the electrons.  The differential cross section can be written as \cite{Harnik:2019zee}
  \be \label{diffSigma}
  \frac{d\sigma}{dE_r}=\frac{e^2 q_C^2 }{8 \pi m_e v^2}\frac{1}{E_r^2} \left( 1+ \frac{2 m_e E_r}{2 m_e E_r+m_{A'}^2}\frac{\delta -\epsilon}{\epsilon}\right)^2  \ee 
  {\rm where} $E_r<E_{max}=2 m_e v^2\left(\frac{m_C}{m_C+m_e}\right)^2$.
  The  first (second) term in the parenthesis comes from the contribution from $t$ channel photon (dark photon, $A'$)
 exchange.  
  
  Analyzing  the XENON1T data, we take $v_f=0.08$, $q_C=10^{-11}$ and $m_X=10$~MeV and   use the data presented  in Fig.~4 of Ref.~\cite{Aprile:2020tmw}. Let us  define $\chi^2$ as follows
 \begin{equation} \label{chi2}
 \chi^2=
 \sum_{bins}\frac{[N^{obs}_i - N^{pred}_i ]^2 }{\sigma_i^2}
 \end{equation}
 where $N^{obs}_i$ is  the observed number of  events at  bin $i$ and $N^{pred}_i$ is the prediction for the sum of the   background and the signal from the $C$ scattering in the $i$th bin. The values of the uncertainty  ($\sigma_i$), the background and  $N^{obs}_i$  are obtained from Fig.~4 of Ref.~\cite{Aprile:2020tmw}.
 We only focus on  the first seven bins because the  excess appears only  at $E_r<8$~keV.  The variation of $\chi^2$  with $m_C$ seems to be  only mild. 
 
\begin{figure}[t]
	%\begin{center}
	\includegraphics[scale=0.85]{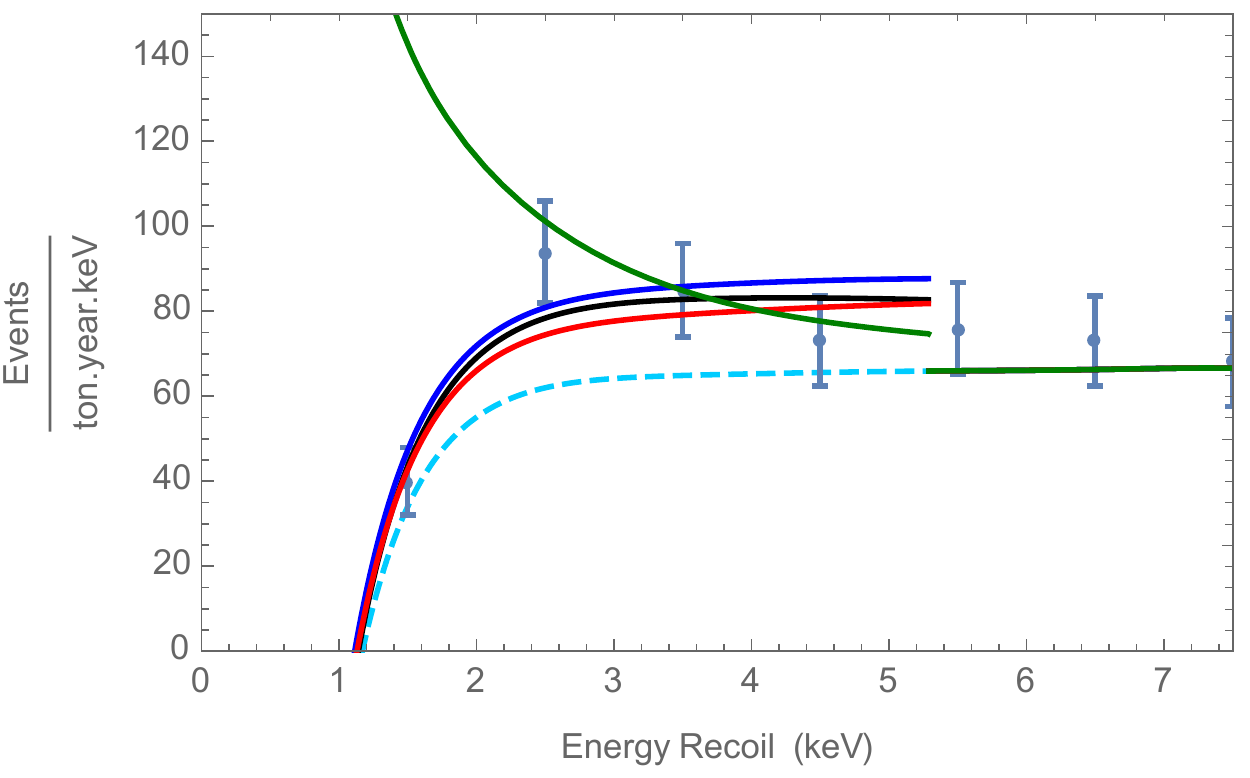}
	\caption{ XENON1T datapoints superimposed on  our predictions. We have set    $m_C= 5$~MeV corresponding  to $E_{max}=5.3$ keV.
		The red   curve  shows  the prediction  for $m_{A^\prime} > 1$ MeV,  $f=10^{-7}$ and $(\delta -\epsilon)/\epsilon=-2 \times 10^3 (m_{A'}/{\rm MeV})^2$.
		The black curve depicts  the prediction for  $m_{A^\prime} = 0.1$ MeV, $(\delta -\epsilon)/\epsilon=-13.9 $ and $ ~ f=6.6\times 10^{-7}$.
		The blue curve illustrates  the prediction for  $m_{A^\prime} = 0.2$ MeV, $(\delta -\epsilon)/\epsilon=-76 $ and $ ~ f=2\times 10^{-7}$. The green curve shows the SM photon dominant regime with $f=5\times 10^{-6}$. The  SM background is shown by the cyan dashed curve. 
	} \label{fit2}
	%\end{center}
\end{figure}

 In sect.~\ref{model}, we discussed that for $100~{\rm keV}<m_{A'}< $few MeV, identifying the $U_X(1)$ with the $L_\mu-L_\tau$ gauge symmetry, the bounds on $q'\propto  (\delta -\epsilon$) from supernova cooling can be relaxed because $A'$ can decay into neutrino pairs relatively fast. 
For  $m_{A'}\sim$few MeV, $2m_e E_r\ll m_{A'}^2$ and  we can treat $(\delta -\epsilon)/(\epsilon m_{A'}^2)$  and $f$  as two free parameters to be fitted with the data. Taking $m_C=5$~MeV, we find that  the minimum of $\chi^2$ lies at $f=10^{-7}$ and $(\delta -\epsilon)/\epsilon=-2 \times 10^3 (m_{A'}/{\rm MeV})^2$. For $\epsilon=-q_C/(g_X\cos \theta_W)\sim 10^{-11}/(g_X\cos \theta_W)$, this corresponds to $q' \sim (10^{-8}/g_X) (m_{A'}/{\rm MeV})^2$. 	At this point, $\chi^2=4.7$ with $5=7-2$ degrees of freedom which corresponds to a one-sided goodness of fit of 45~\%. 
The results are shown in  Fig.~\ref{fit2} with a red curve. In the first bin, contribution from the dark photon and the SM photon cancel each other.
In case of the green curve shown in Fig.~\ref{fit2}, we have taken $2 m_eE_r (\delta /\epsilon-1)/m_{A'}^2\to 0$ so there is no cancellation between dark and SM photons. As seen in the figure, in this case the spectrum diverges for small $E_r$.

%%%%%%%%%%%%%%%%%%%%%%%%
%%%%%%%%%%%%%%%%%%%%%%%%%%%%%%%%
%%%%%%%%%%%%%%%%%%%%%%%%%%%%%%%
%%%%%%%%%%%%%%%%%
%%%%%%%%%%%%%%%%%%%
\begin{figure}[t]
	%\begin{center}
	
	\includegraphics[scale=0.7]{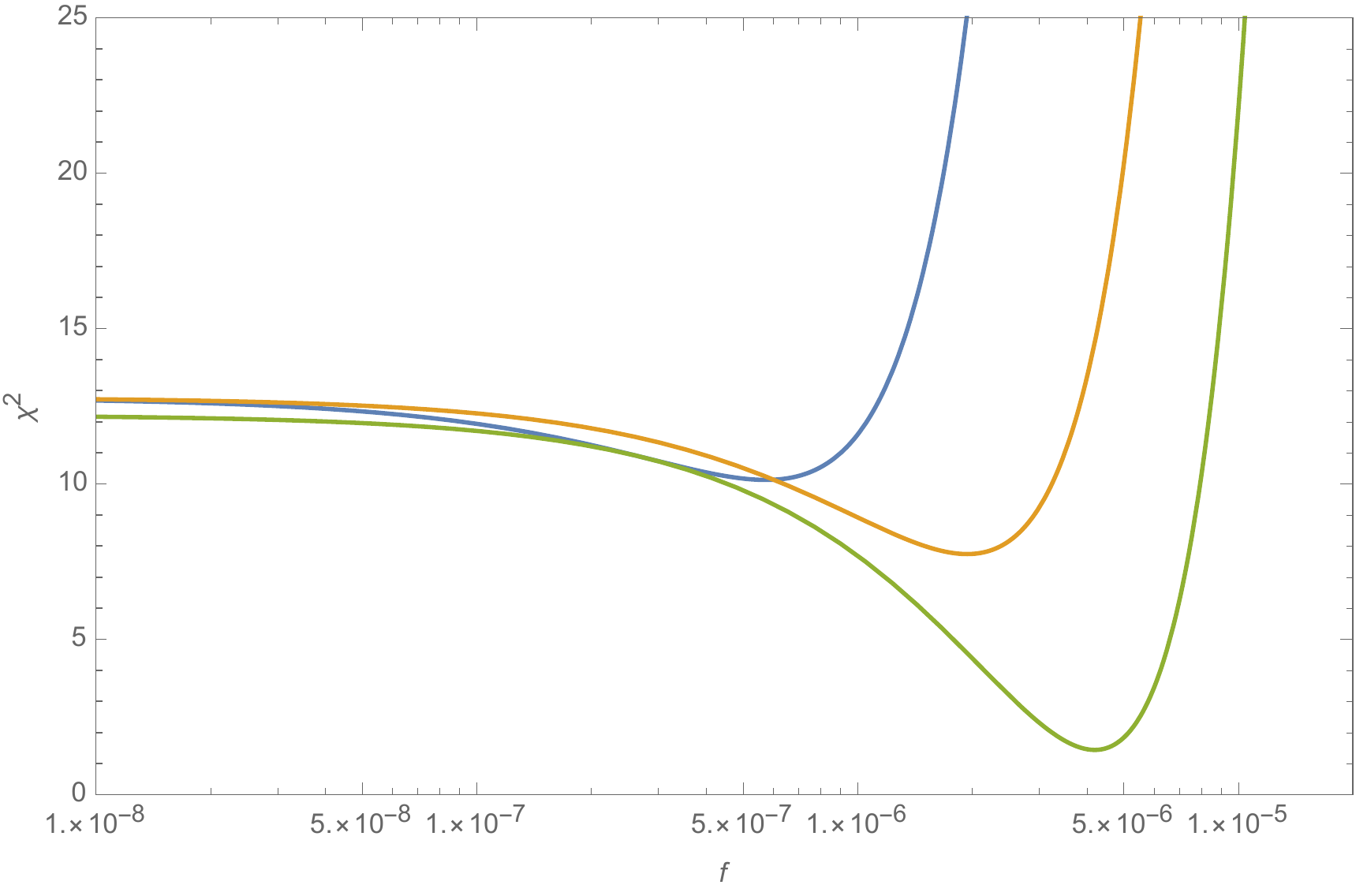}
	\caption{ $\chi^2$ vs. $f$.  
		We have taken $v_f=0.08$, $q_c=10^{-11}$, $m_X=10$~MeV and $m_C=4$~MeV.
		The blue and orange  curves (the green curve) show $\chi^2$ for the seven (six) bins between 1~keV$<E_r<$ 8~keV (2~keV$<E_r<$ 8~keV).  To draw the blue and green curves, we have assumed that the photon exchange gives the dominant contribution to the scattering, $(m_eE_r/m_{A'}^2)[(\delta -\epsilon)/\epsilon]\ll 1$. To draw the orange curve we have set $(2m_e/m_{A'}^2)[(\delta -\epsilon)/\epsilon]=-1/(1.5~{\rm keV})$. } \label{chi}
	%\end{center}
\end{figure}
%%%%%%%%%%%%%%%%%%
%%%%%%%%%%%%%%%%%
For $m_{A'}\sim 100$ keV, $2 m_e E_r$ is comparable to $m_{A'}^2$ and we should use the whole formula to fit the data. 
Taking $m_C=5$~MeV and $m_{A'}=100$~keV ($m_{A'}=200$~keV), the best fit will correspond to
 $(\delta -\epsilon)/\epsilon=-13.9$ and $f=6.6\times 10^{-7}$ ($(\delta -\epsilon)/\epsilon=-76$ and $f=2\times 10^{-7}$) and to the minimum $\chi^2=4.11$ ($\chi^2=4.5$).
 From  Fig.~\ref{fit2}, we observe that   both fits result in very close prediction. 
 At this fit, $q'$ is  right above the supernova bound ($q'\sim 10^{-10}-10^{-9}$) which can be ruled in by opening the possibility of decay of $A'$ into neutrino pairs in the outer layers of the supernova as discussed in sect.~\ref{model}. The effects of such couplings can be tested by studying   the flavor composition, the neutrino energy spectrum and the duration of the neutrino emission. The $q'$ coupling can lead to the production of the  $A'$ particles  in the early universe, contributing to the extra relativistic degrees of freedom,  $\delta N_{eff}<1$. More accurate  $\delta N_{eff}$  determination  by future CMB and BBN studies can test this model 
 with $q'\sim 10^{-10}-10^{-9}$.
 
 Let us now discuss the bound that can be derived on $f$ from the XENON1T data. To do so, we consider the bins with energy less than 8 keV.
Fig.~\ref{chi} shows  $\chi^2$ as a function of  $f$. Since the first bin is close to the detection threshold and it suffers from large uncertainty in the evaluation of the background from $^{214}$Pb, we present results both including and excluding the first bin. We have taken $m_C=4$~MeV, but the results are rather insensitive to the value of $m_C$.  To draw the blue and green lines, we have set $\delta=\epsilon$ which means the dark photon decouples and does not contribute to the scattering. Drawing the orange line, we have chosen the value of $2 m_e(\delta -\epsilon)/(m_{A'}^2 \epsilon)$ such that the contributions from SM photon and dark photon cancel each other at $E_r=1.5$~keV.    From the figure  we find  an upper bound of $1.5 \times 10^{-6}$ on $f$ at 3~$\sigma$,  taking all the bins with $E_r<8$~keV:
$$ f<1.5 \times 10^{-6}   \ \ \ {\rm at}\ 3 \sigma.$$
Excluding the first energy bin  relaxes the bound to $10^{-5}$. Notice that these bounds are much stronger than the constraints discussed in Sect.~\ref{model}.  Since $f=\Gamma_X t_0$, the upper bound on $f$ is equivalent to  a lower bound on the dark matter lifetime. The lifetime of $X$ should be larger than $10^{15}-10^{16}$ years.  

Let us revisit the explanation for the 511 keV line taking into account this very stringent upper bound on $f$.
To explain the intensity  of the 511 keV line from the galactic bulge, the annihilation cross section of $C\bar{C}\to \phi {\bar{\phi}}$ should be ${\rm mb} ( 10^{-7}/f)^2 (m_X/5~{\rm MeV})^2$. 
Taking a quartic coupling between these scalars of form $\lambda_{\phi C}|\phi|^2 |C|^2$, we need  $\lambda_{\phi C}\sim 0.02$ to obtain  an annihilation cross section of mb.
\section{Conclusion and discussion \label{sum}}

We have presented a dark matter model that can simultaneously explain the 511~keV line from the galactic bulge  and the electron excess with recoil energy of $2-4$ keV reported by XENON1T. The model is based on introducing a dark matter particle with a mass of $O(1-10)$ MeV that decays into a pair of pico-charged particles.  Although the velocities of the $C$ and $\bar{C}$ particles are much larger than the escape velocity from the galaxy, the galactic magnetic field can accumulate them inside the galactic disk. The $C$ particles with velocities of order of $0.08c$ can scatter off the electrons of the direct dark matter search experiments, imparting a recoil energy of a few keV which can explain the electron excess at the low energy bins reported by XENON1T. Moreover, the $C\bar{C}$ accumulated in the galactic bulge can annihilate with each other, giving rise to the excess of the 511 keV line. We have shown that the XENON1T  data also sets strong upper bound of $10^{-6}$ on the fraction of $X$ particles that have decayed into $C\bar{C}$ since early universe until today, $f<10^{-6}$. The density of $C\bar{C}$ at the time of recombination and during the dark ages would be too low to pump any significant energy and ionize the atoms so the bounds from the CMB  on  the minimalistic DM explanation for the 511 keV line do not apply here. 
 
 The model enjoys a rich phenomenology and can be tested by a myriad of methods. Most obvious test is further data from upcoming direct dark matter search experiments, especially the XENONnT detector which is an upgrade of XENON1T. The XENONnT detector is in the construction phase.
 
  The $C$ particles obtain their electric charge through the mixing of the photon and the dark photon, $A'$ which is the gauge boson of a new $U_X(1)$ symmetry under which $C$ and $\bar{C}$ are charged.  Simultaneous explanation for  the excess of events with recoil energy between $2-4$ keV and for  the absence of an excess at the 1 keV bin requires a cancellation between the contributions from the $t$-channel exchanges of the photon and $A'$. Such a cancellation is possible with a light $A'$ with a mass of 100 keV--1 MeV. We have shown that by identifying the $U_X(1)$ symmetry with the $L_\mu-L_\tau$ symmetry and therefore  allowing for $A' \to \nu_\mu\bar{\nu}_\mu,  \nu_\tau\bar{\nu}_\tau$, it is possible to avoid the present bounds from the supernova cooling consideration. The $A'$ decaying to $\nu \bar{\nu}$ in the outer layers of supernova can however alter the evolution of supernova,  leaving its imprint on  the duration of neutrino emission, flavor composition, energy spectrum and shock revival. All these signatures provide alternative methods to test the model.
 
 The intensity of the 511 keV line from dwarf galaxies is proportional to the product of the number of $C$ and $\bar{C}$ particles. As a result, there should be a correlation between the magnitude of the magnetic field in dwarf galaxies and the intensity of the 511 keV photon line emitted from them which can be tested in the future.
 
  The positrons from the $C \bar{C}$ annihilation at the production  are relativistic. A fraction of these positrons may annihilate in flight giving rise to a continuous $\gamma$ ray spectrum on which there is  a bound, implying that the $C$ particles should be lighter than $6-15$ MeV depending on the models for ionization fraction in the galaxy \cite{Siegert:2015knp,Beacom:2005qv,Sizun:2006uh}.
 In future with more data, lighter $C$ can be probed by this consideration. Moreover, there will be an excess cosmic positron at few MeV range. At these energies, the  solar wind renders measurements of cosmic positron flux in the  MeV range  obsolete within the solar system. Fortunately, there are detectors on board the Voyager spacecraft  that have set  an upper bound on the cosmic positrons outside  the solar system. The bound can be interpreted as $m_C<20$ MeV. Further data from Voyager can test smaller values of $m_C$ or hopefully detect a positron excess which will be a hint in favor of the present model.
%%%%%%%%%%%%%%%
%%%%%%%%%%%%%%%%

\section*{Acknowledgments}
The article is prepared for the proceedings of the sixteenth Marcel Grossmann meeting (MG16). The author would like to thank the organizers and conveners of this meeting for the kind invitation. She is also grateful for  ICRANET whose support made this participation possible.

This project has received funding/support from the European Union’s Horizon 2020 research and innovation programme under the Marie Skłodowska--Curie grant agreement No 860881-HIDDeN. The author has also received partial financial support from Saramadan under contract No.~ISEF/M/98223, No.~ISEF/M/9916
and No.~ISEF/M/400188.

%Non BiBTeX users can list down their references as:

\end{document}